\documentclass[floatfix,prl,english, twocolumn,superscriptaddress]{revtex4-1}
\usepackage{graphicx,amsfonts,amssymb,amsmath,
	enumerate,upgreek,tabularx,multirow,dsfont,cancel}
\usepackage[displaymath,mathlines]{lineno}
\usepackage[dvipsnames]{xcolor}
\definecolor{darkblue}{rgb}{0.0, 0.0, 0.75}
\usepackage{color}
\usepackage[colorlinks=true,
linkcolor=darkblue,
urlcolor=darkblue,
citecolor=darkblue]{hyperref}
\newcommand{\subfig}[2]{%
	{\textsf{#1}} \vtop{%
		\vskip0pt
		\hbox{#2}
}}

\newcommand{\comment}[1]{}

\newcommand{\llangle}{\langle\hspace{-0.25em}\langle}
\newcommand{\rrangle}{\rangle\hspace{-0.25em}\rangle}

\newcommand{\ket}[1]{\mathopen{\mid}#1\mathclose{\rangle}}

\newcommand{\brav}[1]{\mathopen{\llangle}#1\mathclose{\mid}}
\newcommand{\ketv}[1]{\mathopen{\mid}#1\mathclose{\rrangle}}

\newcommand{\braket}[1]{\langle #1\rangle}

\newcommand{\braketv}[1]{\brav{#1} #1 \mathclose{\rrangle}}
\newcommand{\braketvp}[2]{\brav{#1} #2 \mathclose{\rrangle}}
\newcommand{\braketvpp}[2]{\brav{#1} #2 \ketv{#1}}
\newcommand{\norm}[1]{\left\lVert#1\right\rVert}
\begin{document}

\title{Driven-dissipative Rydberg blockade in optical lattices}

\author{Javad Kazemi}
\email{javad.kazemi@itp.uni-hannover.de}
\affiliation{Institut f\"ur Theoretische Physik, Leibniz Universit\"at Hannover, Appelstra{\ss}e 2, 30167 Hannover, Germany}
\author{Hendrik Weimer}
\email{hweimer@itp.uni-hannover.de}
\affiliation{Institut f\"ur Theoretische Physik, Leibniz Universit\"at Hannover, Appelstra{\ss}e 2, 30167 Hannover, Germany}

\begin{abstract}

  While dissipative Rydberg gases exhibit unique possibilities to tune
  dissipation and interaction properties, very little is known about
  the quantum many-body physics of such long-range interacting open
  quantum systems. We theoretically analyze the steady state of a van
  der Waals interacting Rydberg gas in an optical lattice based on a
  variational treatment that also includes long-range correlations
  necessary to describe the physics of the Rydberg blockade, i.e., the
  inhibition of neighboring Rydberg excitations by strong
  interactions. In contrast to the ground state phase diagram, we find
  that the steady state undergoes a single first order phase
  transition from a blockaded Rydberg gas to a facilitation phase
  where the blockade is lifted. The first order line terminates in a
  critical point when including sufficiently strong dephasing,
  enabling a highly promising route to study dissipative criticality
  in these systems. In some regimes, we also find good quantitative
  agreement with effective short-range models despite the presence of
  the Rydberg blockade, justifying the wide use of such
  phenomenological descriptions in the literature.

\end{abstract}
  
\maketitle


Strongly interacting Rydberg atoms undergoing driving and dissipation
allow to stud a wide range of many-body effects not seen in their
equilibrium counterparts, ranging from dissipative quantum sensors
\cite{Wade2017} to self-organizing dynamics \cite{Helmrich2020}. These
effects are enabled by the presence of a strong van der Waals
interaction between the atoms, which is fundamentally
long-ranged. Yet, very little is known about the many-body properties
of such systems, as long-ranged open quantum systems are inherently
hard to simulate on classical computers \cite{Weimer2021}. Here, we
present a variational calculation of the steady state phase diagram of
a long-range interacting Rydberg in an optical lattice.

A key feature of strongly interacting Rydberg gases is the appearance
of the Rydberg blockade, where the strong van der Waals interactions
prevents the excitation of neighboring Rydberg atoms
\cite{Jaksch2000,Lukin2001,Singer2004,Heidemann2007,Schauss2012,Bernien2017}. However,
given the intrinsic challenges associated with the treatment of open
quantum many-body systems, most works studying dissipative Rydberg
gases have employed short-range interactions
\cite{Lee2012,Weimer2015,Weimer2015a,Kshetrimayum2017,Panas2019,Singh2022,Kazemi2021b},
which are inadequate for strongly blockaded Rydberg gases. As a
consequence, the steady state phase diagram of blockaded Rydberg gases
is essentially unknown.

In this Letter, we present the application of a variational approach
for the non-equilibrium steady state of Rydberg atoms with strong
repulsive van der Waals interactions. Notably, we explicitly account
for correlations between multiple Rydberg excitations. Based upon the
variational results, we find a dissipative variant of the Rydberg
blockade in the pair correlation function, where any simultaneous
excitation within a certain blockade radius is strongly suppressed. In
addition, we investigate the interplay between coherent driving and
dissipation, finding a first-order dissipative phase transition, which
terminates in a critical point under sufficiently strong additional
dephasing. Finally, we analyze the validity of effective short-range
models to describe the dynamics even in the blockaded regime, where we
find that such effective models actually perform better in the
presence of a strong Rydberg blockade.

\begin{figure}[t]
	\begin{center}
	  \begin{tabular}{cc}	
		  \subfig{a}{\includegraphics[height=3.7cm]{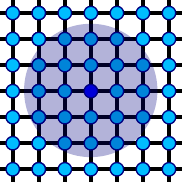}} &
          \subfig{b}{\includegraphics[height=3.7cm]{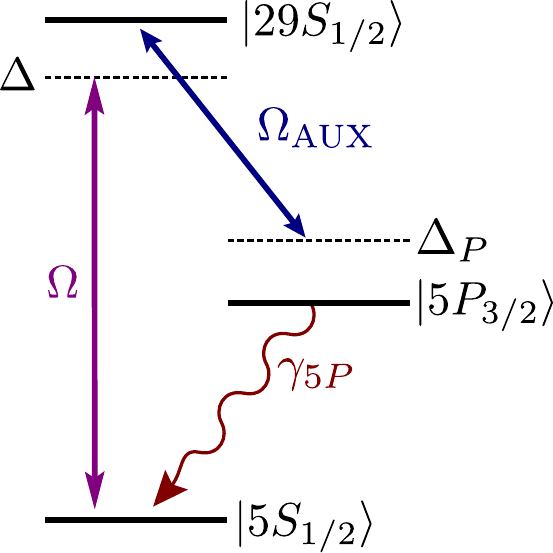}}
          \end{tabular}
        \end{center}
        \caption{Setup of the system. (a) Strong van der Waals
          repulsion leads to a blockade of simultaneous Rydberg
          excitations over a number of sites of the square
          lattice. (b) Internal level structure showing a two-photon
          excitation of a single Rydberg state with a Rabi frequency
          $\Omega$ and a detuning $\Delta$, as well as laser-assisted
          dissipation via an intermediate excited state.}
        \label{fig:setup}
\end{figure}

\emph{Model.---} Dissipative processes involving electronic
excitations can often be described in terms of a Markovian master
equation in Lindblad form as the frequency of the emitted photons
provides for a natural separation of timescales required for the
Markov approximation \cite{Breuer2002}. Then, the time evolution of
the density matrix $\rho$ is given by
\begin{equation}\label{eq:Lindblad}
\mathcal{L}(\rho)= -i [H,\rho] + \sum\limits_{j}\Big[ {c_j} \rho c_j^\dagger - \frac{1}{2} \big\{ c_j^\dagger c_j, \rho \big\}\Big],
\end{equation}
where $\mathcal{L}$ denotes the Liouvillian superoperator $\mathcal{L}(\rho)=\partial_t\rho$, which consists of a coherent part represented by the Hamiltonian $H$ and a dissipative part characterised a set of jump operators $c_{j}$. The coherent dynamics of a ground state atom being laser driven to a single Rydberg state can be expressed in a spin $1/2$ Hamiltonian as
\begin{equation}\label{eq:H_Ryd2}
    H = -\frac{\hbar\Delta}{2} \sum\limits_{i}^N{\sigma_z^{(i)}} + \frac{\hbar\Omega}{2} \sum\limits_{i}^N{\sigma_x^{(i)}} + C_6 \sum\limits_{i<j}{\frac{P_r^{(i)}P_r^{(j)}}{r_{ij}^6}},
\end{equation}
where $\Omega$ is the driving strength, $\Delta$ is the detuning from
the atomic resonance, and $C_6$ denotes the strength of the repulsive
van der Waals interaction \cite{Robicheaux2005,Low2012} involving the
projector $P_r^{(i)}$ onto the Rydberg state. In our work, we consider
the atoms being loaded in a two-dimensional optical lattice, see
Fig.~\ref{fig:setup}. The derivation of the jump operators $c_j$ is
more subtle \cite{Nill2022}, but laser-assisted dissipation via an
intermediate electronic excitation allows to use effective jump
operators of the form $c_j = \sqrt{\gamma}\sigma_-^{(j)}$, with
$\gamma$ being the effective decay rate from the Rydberg state into
the ground state \footnote{See the Supplemental Material for a
  derivation of the effectively local jump operators, the details of
  the variational ansatz, and the derivation of the effective
  short-range model.}. Furthermore, this setup has the advantage that
it allows to tune the dissipation rate independently from the other
properties of the Rydberg state such as the $C_6$ coefficient. Here,
we also assume that the laser-mediated decay is much faster than the
natural decay of the Rydberg excitation or changes in the Rydberg
state by blackbody radiation, therefore we neglect these processes. In
addition to the dissipation, we also allow for a second set of jump
operators $c_j' = \sqrt{\gamma_p}\sigma_z^{(j)}$, where $\gamma_p$
denotes a dephasing rate arising, e.g., from noise of the driving
laser.

Concerning the experimental
setup, we consider a square optical lattice trapping Rubidium-87 atoms
with a lattice spacing of $a=532\,\text{nm}$. Here, we consider
laser driving from the electronic ground state by a two-photon
transition to the state $\ket{29\text{S}_{1/2}}$, which has a van der
Waals coefficient of $C_6 = h \times 17\,\textrm{MHz}\,\mu m^6$
\cite{Weber2017}.

\emph{Variational treatment of long-range correlations.---} In the
following, we will be interested in the properties of the
non-equilibrium steady state given by the condition
$\partial_t\rho=0$. Following the variational principle for steady
states of open quantum systems \cite{Weimer2015}, we turn to a
variational parametrization that also allows for long-range
correlations, which are crucial to capture the physics of the Rydberg
blockade. For the trial state, we consider a variational ansatz
containing local density matrices as well as two-body correlations in
the following form
\begin{equation}
\rho_{\text{var}}=\prod_{i=1}^N \rho_i+\sum_{i<j}\mathcal{R}C_{ij},
\end{equation}
where $N$ is the number of sites, $\mathcal{R}$ is a superoperator transforming the identity matrix $\mathds{I}_i$ into $\rho_i$, and $C_{ij} = \rho_{ij}-\rho_i\otimes\rho_j$ denotes the two-particle correlations. Crucially, we ignore higher-order correlations as in dissipative dynamics with power-law interactions they decay faster than two-body correlations \cite{Navez2010}. Considering the residual dynamics of the ansatz, we define a variational cost function that can be cast into the form
\begin{equation}\label{eq:v_norm}
	F_v \equiv N^{-1}\frac{\norm{\dot{\rho}_{\text{var}}}_\text{HS}^2}{\norm{\rho_{\text{var}}}_\text{HS}^2},
\end{equation}
where $\norm{O}_\text{HS}=\sqrt{\text{Tr}\left[OO^\dagger\right]}$ is the Hilbert-Schmidt norm.
Using the definition of $F_v$ for a transnationally invariant system, and expanding the Liouvillian in terms of local and interacting terms, i.e. $\mathcal{L}=\sum_i \mathcal{L}_{i}+\sum_{i<j} \mathcal{L}_{ij}$, we end up with an efficiently computable upper bound, i.e. $F_v \leq f_v$, that reads as 
\begin{equation}
\begin{split}
f_v&=
g_p^{-2}\sum_{1\neq j}\braketvpp{\rho_{\text{var}}^{(1j)}}{\mathcal{L}_{1}^\dagger \mathcal{L}_{1}+\mathcal{L}_{1}^\dagger \mathcal{L}_{j}}\\
&+g_p^{-3}\sum_{1\neq j\neq k}\braketvpp{\rho_{\text{var}}^{(1jk)}}{2\mathcal{L}_{1}^\dagger \mathcal{L}_{1j}+\mathcal{L}_{1}^\dagger \mathcal{L}_{jk}}\\
&+g_p^{-4}\sum_{1\neq j\neq k\neq l}\braketvpp{\rho_{\text{var}}^{(1jkl)}}{\frac{1}{2}\mathcal{L}_{1j}^\dagger \mathcal{L}_{1j}+\frac{1}{2}\mathcal{L}_{1j}^\dagger \mathcal{L}_{jk}\\
&\hspace{2.7cm}+\frac{1}{2}\mathcal{L}_{1j}^\dagger \mathcal{L}_{1k}+\frac{1}{4}\mathcal{L}_{1j}^\dagger \mathcal{L}_{kl}},
\end{split}
\end{equation}
where we used the notation $\norm{\dot{\rho}_{\text{var}}}_\text{HS}^2=
\braketvp{\mathcal{L}^\dagger(\rho_{\text{var}})}{ \mathcal{L}(\rho_{\text{var}})}$, and $g_p=\braketv{\rho_1}$ denotes the local purity \cite{Note1}.

Within this approach, we can also make statements about systems in the
thermodynamic limit, even for finite $N$. This is possible, as within
our variational approach, a system of $N$ sites is indistinguishable
from an infinitely large system in which correlations over a cluster
of $N$ sites are accounted for. Hence, we can easily detect the
presence of dissipative phase transitions by observing non-analytic
behavior of steady state observables. Note that while our approach
shares some similarities to previous approaches based on cluster
mean-field theory \cite{Jin2016,Jin2018}, the variational character
allows us to avoid some pitfalls associated with mean-field theory in
open systems \cite{Weimer2015,Maghrebi2016,Overbeck2017}.

A crucial aspect of our variational approach is to accurately describe
the pair correlation function
$g_2(r_{ij})=\braket{P_r^{(i)}P_r^{(j)}}/\braket{P_r^{(i)}}\braket{P_r^{(j)}}$
with relatively few variational parameters that need to be
optimized. From perturbation theory in the strongly blockaded regime
\cite{Muller2009}, one can expect that the pair correlation function
behaves for small distances as $g_2(r_{ij})\sim r_{ij}^6$, while for
weak interactions at large interactions, one would expect
$\lim_{r_{ij}\to \infty} g_2(r_{ij}) = 1$. An expansion satisfying these two limits is given by
\begin{equation}
  g_2(r)  = \frac{\alpha_6 r^6}{\alpha_6 r^6 + \sum\limits_{0\leq k < 6} \alpha_k r^k},
  \label{eq:g2}
\end{equation}
where the $\alpha_k$ are variational parameters. Here, we truncate the
sum after the second order, for which we find that Eq.~(\ref{eq:g2})
is in good quantitative agreement with exact numerical simulations of
small systems. For correlation functions involving $\sigma_x$ or
$\sigma_y$, we assume an exponential decay, with the only exception
being the correlator
$\langle\sigma_y^{(i)}\sigma_z^{(j)}\rangle-\langle\sigma_y^{(i)}\rangle\langle\sigma_z^{(j)}\rangle$,
which is inherently linked to $g_2(r)$ through the Lindblad master
equation \cite{Note1}.

\begin{figure*}[t]
	\begin{center}
		\begin{tabular}{ccc}	
			\subfig{a}{\includegraphics[width=.6\columnwidth]
				{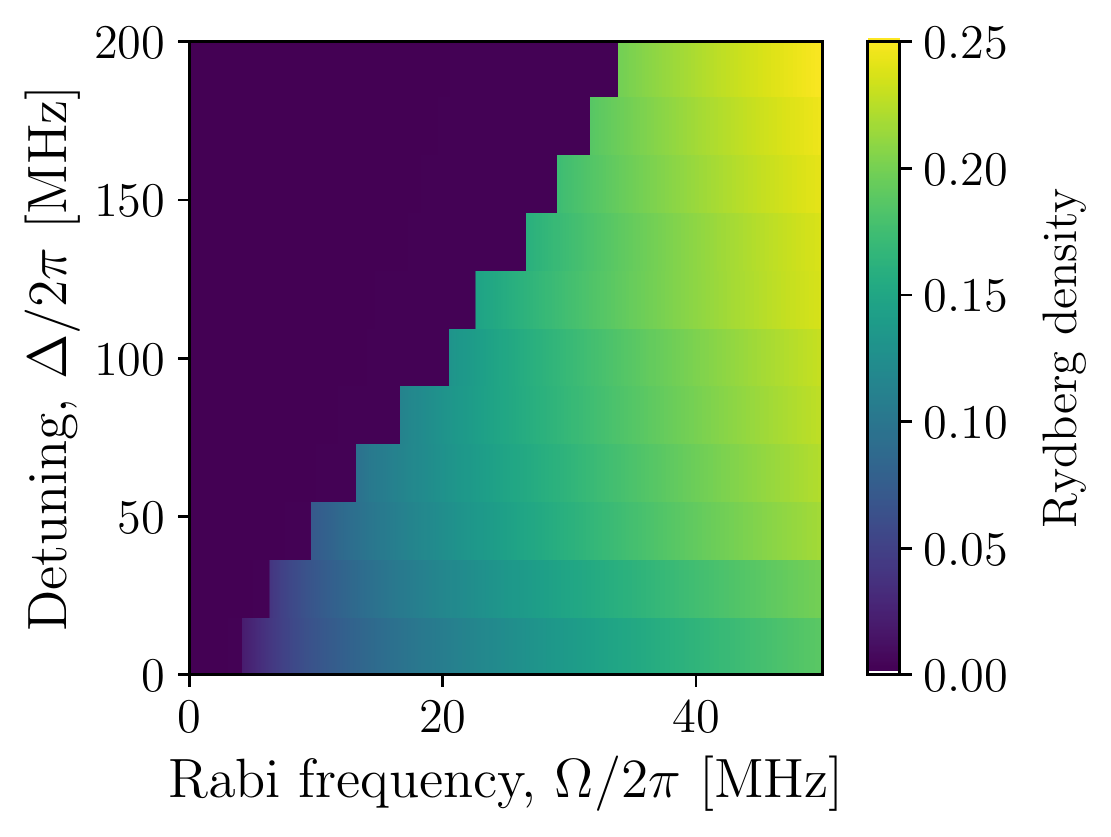}}&\hspace{0mm}
			\subfig{b}{\includegraphics[width=.6\columnwidth]
				{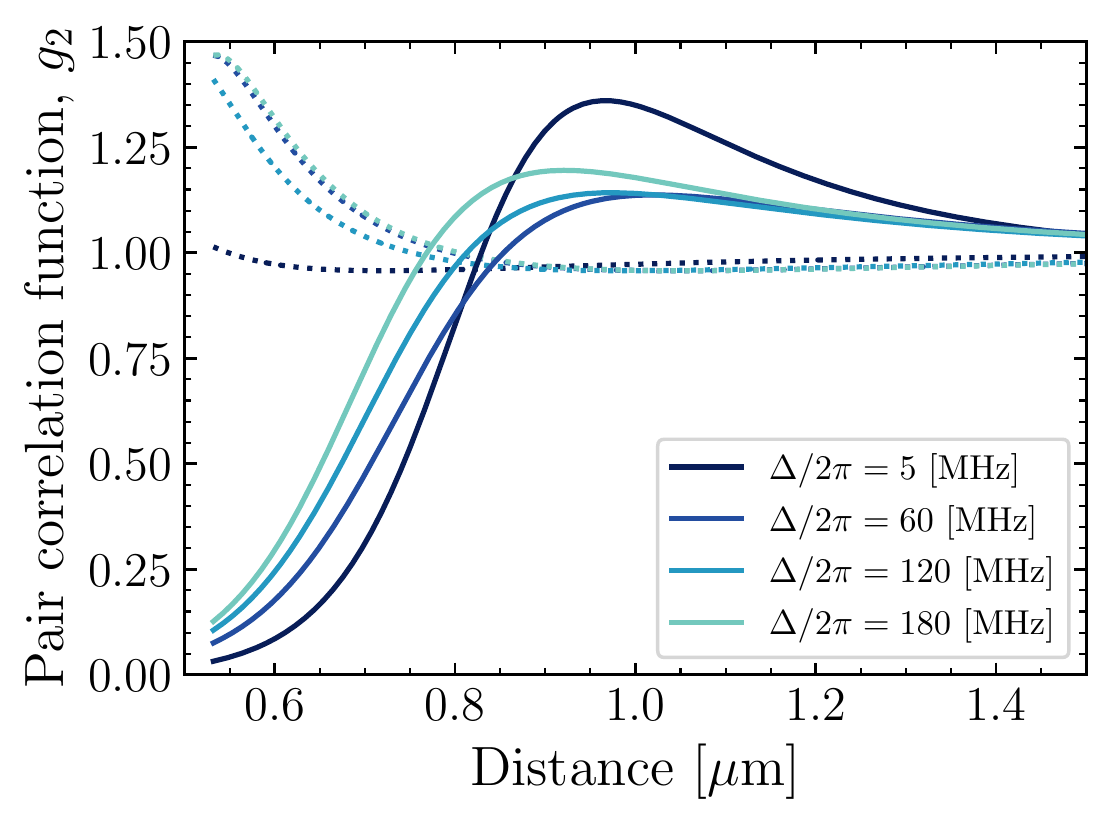}}&\hspace{0mm}
			\subfig{c}{\includegraphics[width=.6\columnwidth]
				{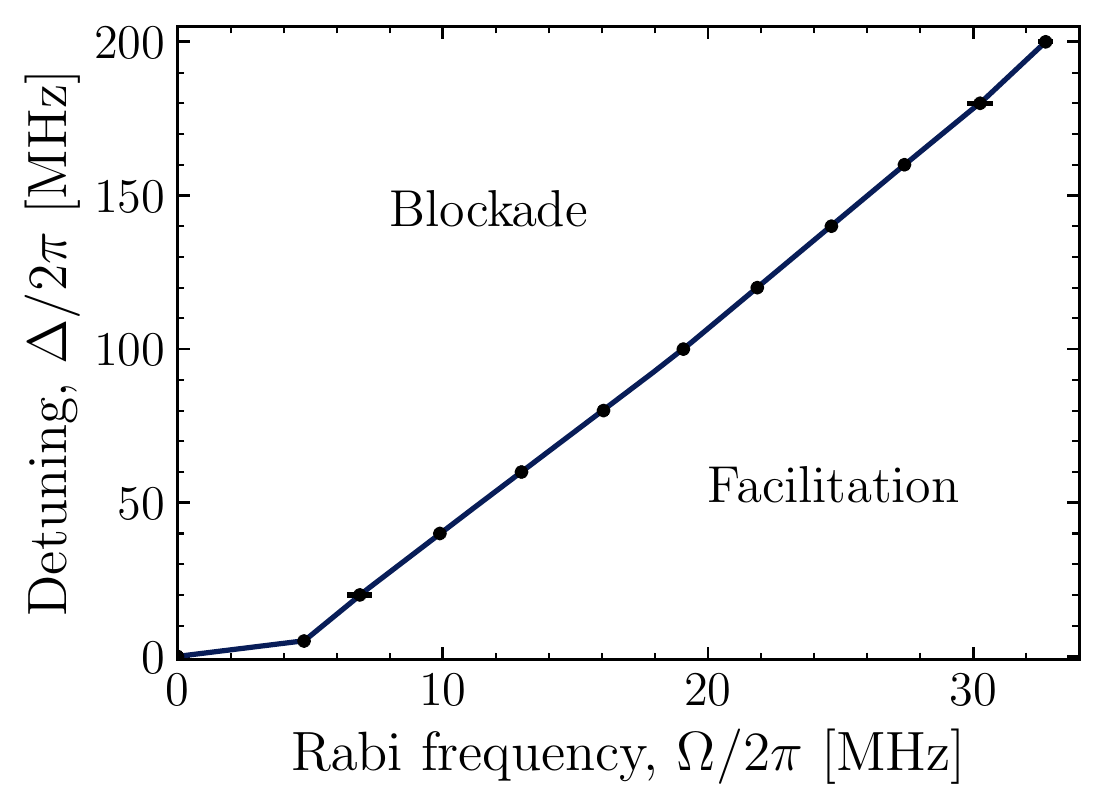}}
		\end{tabular}
	\end{center}
	\caption{Variational results for the steady state of two-dimensional driven Rydberg gases in the presence of decay with the rate $\gamma = 2\pi \times 1\,\text{MHz}$. (a). Rydberg density obtained from a cluster size of $9\times9$ exhibits a discontinuity for different values of detuning $\Delta$ signalling a first-order transition, with the step-like discreteness being due to the numerical resolution. (b) The pair correlation function $g_2$ demonstrates a blockaded region for low-density phase (solid lines) and a facilitated region for high-density phase (dotted lines). In all cases, the driving strength $\Omega$ was chosen right below the transition (solid) or right above it (dotted) (c) The steady-state phase diagram is obtained by finite-size scaling of the variational results which resembles a lattice liquid-gas transition. }
	\label{fig:phase}	
\end{figure*}

\emph{Steady state phase diagram.---} First, we focus on the case without any dephasing,
i.e., $\gamma_p = 0$. Without any dissipation, the ground state phase diagram of
Eq.~(\ref{eq:H_Ryd2}) on a lattice for $\Delta > 0$ consists of a
series of crystalline phases that can be either commensurate or
incommensurate with the underlying lattice, and a paramagnetic phase
for sufficiently strong driving strength $\Omega$
\cite{Weimer2010a,Capogrosso-Sansone2010,Sela2011}. Including
dissipation, we find a vastly different phase diagram for the
nonequilibrium steady state. Overall, we find the presence of two
competing phases, see Fig.~\ref{fig:phase}, one at low densities
exhibiting the characteristic pair correlation function $g_2(r)$
of the Rydberg blockade, and one at higher densities in which the
Rydberg atoms are either uncorrelated or even exhibiting bunching of
neighboring Rydberg excitations, i.e., leading to a complete lifting
of the Rydberg blockade. We note that the second phase continuously
connects to the so-called anti-blockade or facilitated regime
\cite{Ates2007,Urvoy2015,Valado2016}, where the detuning cancels the
longitudinal field term arising from the interaction, resulting in an
effectively purely transversal Ising model. Hence, we refer to the to
phases as ``blockade phase'' and ``facilitation phase'',
respectively. We find that the two phases are separated by a first
order transition that stretches throughout the entire parameter
space. To determine the position of the first-order line in the limit
of infinitely large clusters, we perform an analysis in close analogy
to finite-size scaling, using cluster sizes of $5\times5$, $7\times7$,
and $9\times9$. The choice of odd numbers is twofold; First, it
considerably simplifies the expressions for the variational norm as it
can be constructed around a site at the center of the lattice. Second,
even and odd sites have slightly different scaling behavior, so
focusing on only one of them achieves faster convergence. The result
of the cluster scaling analysis is shown in
Fig.~\ref{fig:phase}c. Furthermore, even in the facilitation phase, we
find that the density of Rydberg excitations is comparatively low
close to the transition, which we attribute to the first excitation
still being suppressed by the van der Waals interaction, which is much
stronger than the driving strength $\Omega$. Remarkably, in stark
contrast to the ground state phase diagram, we find no evidence for
consumerability effects with the underlying lattice, as the first
order line appears to be a smooth function, with other quantities such
as the blockade radius behaving in a similar way.

\begin{figure*}[t]
	\begin{center}
	  \begin{tabular}{ccc}
            \subfig{a}{\includegraphics[width=.64\columnwidth]{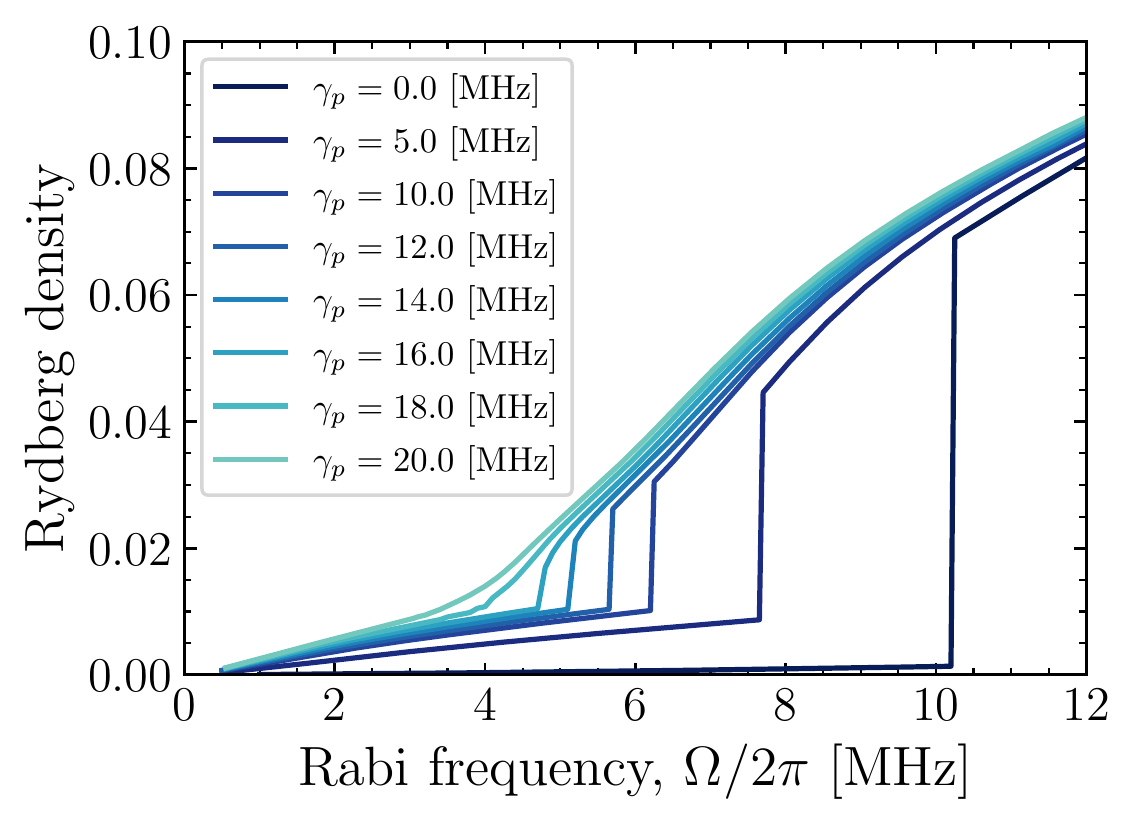}}
            &
            \subfig{b}{\includegraphics[width=.64\columnwidth]{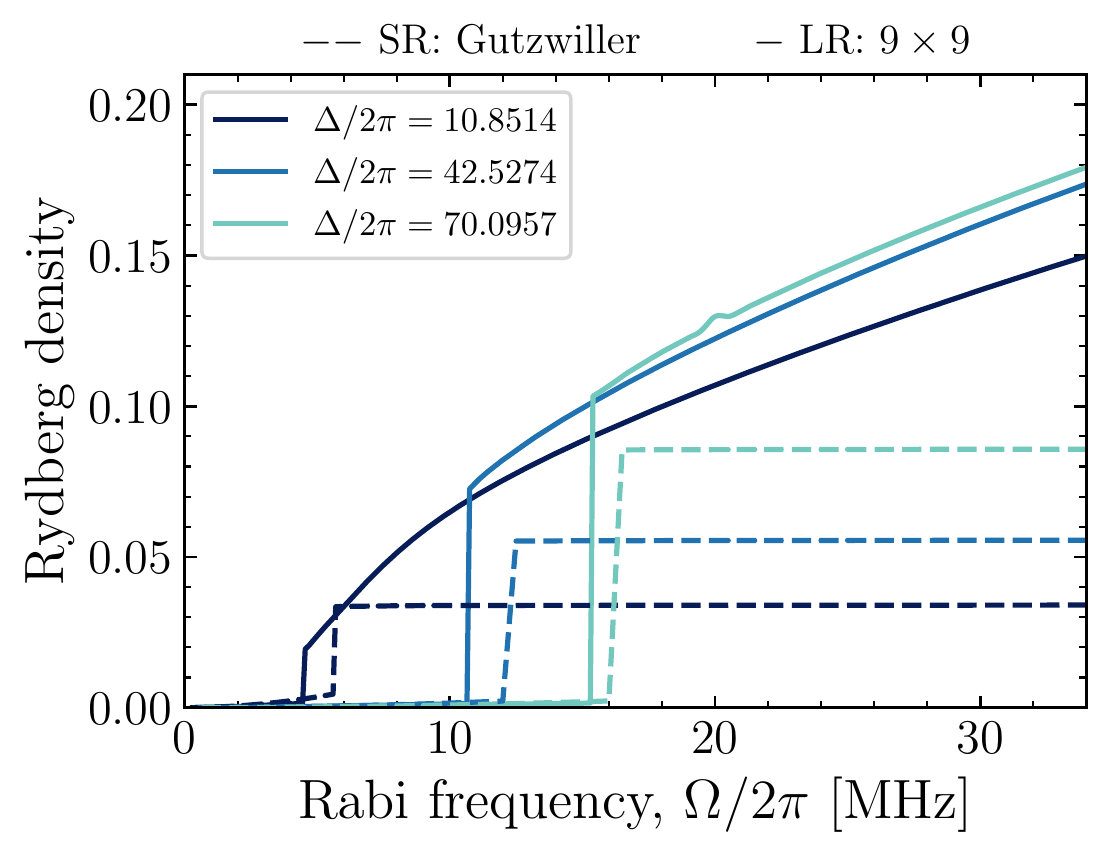}}
            &
            \subfig{c}{\includegraphics[width=.64\columnwidth]{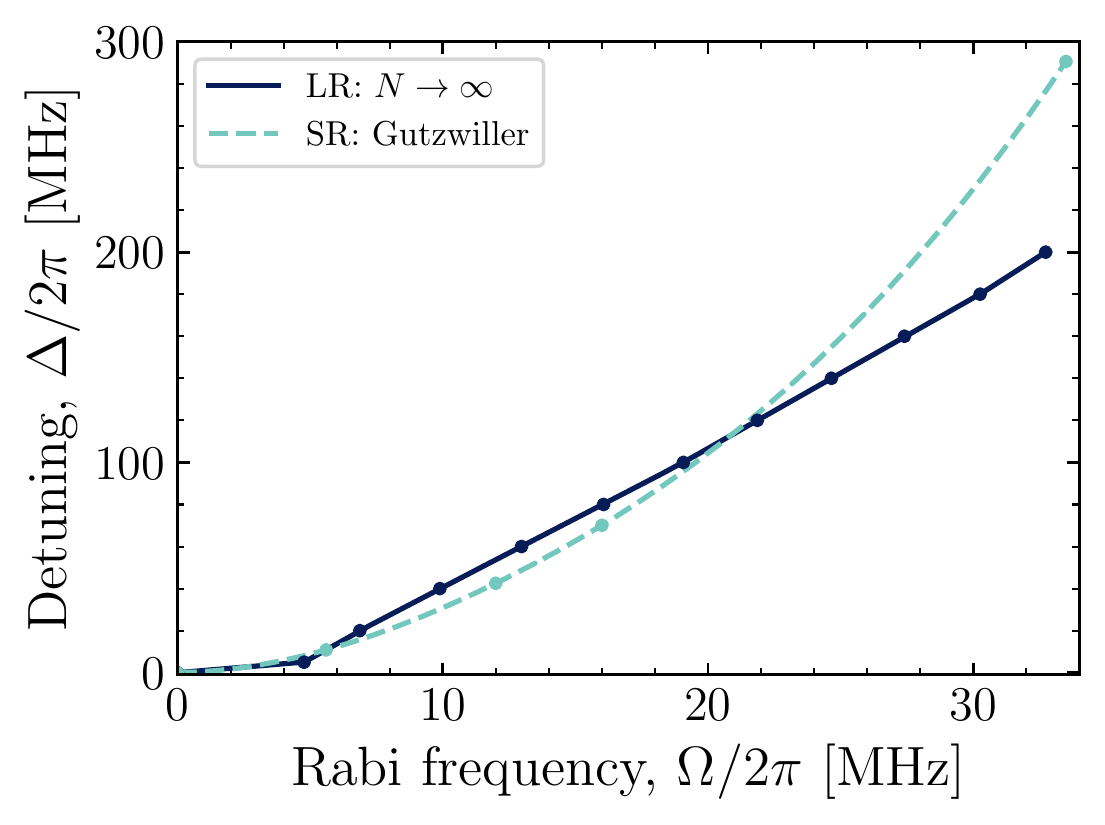}}
		\end{tabular}
	\end{center}
	\caption{Extensions to the basic model. (a) Inclusion of
          dephasing results in the termination of the first order line
          with a critical point. For $\Delta = 2\pi\times
          40\,\textrm{MHz}$, criticality appears at $\gamma_p =
          2\pi\times 16\,\textrm{MHz}$. (b) Variational results for an
          effective short-range interacting model using a Gutzwiller
          ansatz (dashed lines) in comparison with the long-range
          interacting model using the correlated ansatz (solid
          lines). (c) For relatively weak driving, the phase
          transition in dissipative Rydberg gases can be effectively
          described by a short-range interacting model.}
	\label{fig:extension}	
\end{figure*}

\emph{Dephasing-induced criticality.---} Without dephasing, the
numerical results are consistent with the first order line extending
all the way to $\Delta = 0$, without triggering any critical behavior,
in contrast to both the ground state phase diagram and the
steady-state phase diagram for purely coherent evolution
\cite{Weimer2008a,Low2009}. However, as already known from other
dissipative models exhibiting first-order transitions
\cite{Marcuzzi2014,Weimer2015a,Raghunandan2018}, such first order
lines between two competing phases can be turned critical by injecting
additional dephasing noise into the dynamics. Within our variational
approach, we also observe this behavior for the long-range interacting
case, see Fig.~\ref{fig:extension}a. In particular, we find the critical
point emerging for a detuning of $\Delta = 2\pi\times
40\,\textrm{MHz}$ at $\gamma_p = 2\pi\times 16\,\textrm{MHz}$. Such a
level dephasing can be realized experimentally by noise on the
excitation laser, opening a very promising route to study dissipative
criticality in such a setting.

\emph{Effective short range models.---} As already mentioned, the
inherent challenges with studying open quantum many-body stystems has
led to most theoretical works replacing the long-range interaction in
dissipative Rydberg gases by a short-range interaction. While it is
clear that this is inadequate to correctly describe the Rydberg
blockade, one may ask whether such a simplification can still be used to
describe dissipative Rydberg gases, especially since properly
renormalized short-range models have had some success in describing
experiments even in the blockaded regime \cite{Weimer2015a}. To this
end, we develop a systematic way to derive effective short-range
models \cite{Note1}. Here, we consider the situation where the
detuning is chosen in such a way that it realizes an anti-blockade
configuration at a particular distance $r'$. Then, we assume that all
lattice sites in between are effectively frozen and consider an
effective model involving only the remaining sites. As we operate in an anti-blockade configuration, we obtain a purely transversal Ising model, i.e.,
\begin{equation}
H_{\text{eff}} = \frac{\hbar\Omega}{2} \sum_{i}{\sigma_x^{(i)}} + J_{\text{eff}}(r_d) \sum_{\braket{ij}}\sigma_z^{(i)}\sigma_z^{(j)},
\end{equation}
where $J_{\text{eff}}(r_d)$ denotes the strength of the
nearest-neighbor Ising interaction which accounts for van der Waals
interaction excluding terms from the nearest-neighbor up to a
truncation distance $r_d<r'$ \cite{Note1}.

We can benchmark the validity of the effective short-range model by
performing variational calculations. For the short-range model, we
utilize a Gutzwiller ansatz, which has been shown to give reliable
quantitative estimates of the position of the first order transition
\cite{Kazemi2021b}. Fig.~\ref{fig:extension}b-c confirms that the
short-range Ising model is in good quantitative agreement with the
long-range interacting model after accounting for the number of
excluded sites by rescaling the Rydberg density by a factor of
$(r_d+1)^2$, provided that the driving strength $\Omega$ is relatively
small. For larger values of the driving strength, resulting in a
decreasing truncation radius $r_d$, the correspondence becomes worse,
although the position of the first order transition remains quite
accurate over a substantially larger region. This is somewhat
counterintuitive, as in the limit where the truncation distance
vanishes, one would expect a short-range model to be valid, although
there are cases when considering classical rate equations instead of a
Lindblad master equation, where a van der Waals interactions is not
decaying sufficiently fast to allow for a replacement by a
nearest-neighbor interaction \cite{Hoening2014}.

In summary, we have analyzed the stationary properties of strongly
interacting Rydberg gases under driving and dissipation using a
variational treatment properly capturing long-range correlations. In
stark contrast to the ground state phase diagram, the steady state of
dissipative Rydberg gases features a single first-order phase
transition between a blockaded phase and a facilitation phase, without
any commensurability effects from the underlying optical lattice. The
first-order line can be tuned critical by incorporating additional
dephasing noise, leading to a very promising route to investigate
dissipative criticality in an open quantum system. Finally, we find
that despite strong blockade effects, effective short-range
descriptions of dissipative Rydberg gases can be successfully used to
explain the basic features of the steady state phase diagram and hence
can serve as a minimal model.

\begin{acknowledgments}
  We thank C.~Gro{\ss}, P.~Schau{\ss}, and J.~Zeiher for valuable
  discussions on the experimental realization of dissipative Rydberg
  gases. This work was funded by the Volkswagen Foundation, by the
  Deutsche Forschungsgemeinschaft (DFG,German Research Foundation)
  within SFB 1227 (DQ-mat, project A04), SPP 1929 (GiRyd), and under
  Germanys Excellence Strategy -- EXC-2123 QuantumFrontiers -- 90837967.
\end{acknowledgments}

\end{document}


\title{Supplemental Material for ``Driven-Dissipative Rydberg Blockade in Optical Lattices''}
	
	\author{Javad Kazemi}
	\email{javad.kazemi@itp.uni-hannover.de}
	\affiliation{Institut f\"ur Theoretische Physik, Leibniz Universit\"at Hannover, Appelstra{\ss}e 2, 30167 Hannover, Germany}
	\author{Hendrik Weimer}
	\email{hweimer@itp.uni-hannover.de}
	\affiliation{Institut f\"ur Theoretische Physik, Leibniz Universit\"at Hannover, Appelstra{\ss}e 2, 30167 Hannover, Germany}
	\maketitle

        \section{Effective local jump operators for interacting Rydberg atoms}
        \label{SM:local_jumps}

        In many phenomenological descriptions of dissipative Rydberg
        gases, the dissipation is incorporated in terms of local jump
        operators connecting the Rydberg state to the ground
        state. While this is a reasonable assumption for
        non-interacting atoms, the presence of strong van der Waals
        interactions can modify this picture, as the frequency of
        emitted photons generically contains information about the
        local surroundings of the emission site
        \cite{Nill2022}. However, when using a setup involving laser-assisted dissipation, see Fig.~1b of the main text, one can justify the use of local jump operators, as we show below.

        Without loss of generality, we consider a subsystem of three
        Rydberg atoms with three electronic levels for each atom,
        $\{\ket{g_i},\ket{p_i},\ket{r_i}\}$. For simplicity, we also
        neglect the laser driving between the ground state and the
        Rydberg state, as this occurs on a much slower
        timescale. Then, the Hamiltonian is of the form
\begin{equation}
H_{\text{Ryd}} = \Omega_{\text{AUX}} \sum_{i=1}^{3}\left(\ket{p_i}\bra{r_i} + \ket{r_i}\bra{p_i} \right)
+ \Delta_p \sum_{i=1}^{3}\ket{p_i}\bra{p_i}
+ V_{rr} \sum_{\braket{ij}}\ket{r_i r_j}\bra{r_i r_j},
\end{equation}
where $\braket{ij} \in \{12,23\}$  due to the assumption of open boundary conditions. We also consider a spontaneously decay from the intermediate state $\ket{p_i}$ to the ground state $\ket{g_i}$, i.e.
\begin{equation}
c_{i} = \sqrt{\gamma_p} \ket{g_i}\bra{p_i},
\end{equation}
where $\gamma_p$ denotes the decay rate. In the following, we apply
the effective operator formalism \cite{Reiter2012} to eliminate the
intermediate states and obtain effective jump operators only involving the ground and Rydberg states. For this, we divide the Hilbert space of our model into a ground-state subspace $\mathcal{H}_{gr}$, and an excited-state subspace $\mathcal{H}_{e}$. Accordingly, one can decompose the Hamiltonian as follows
\begin{equation}
\begin{split}
H_{\text{Ryd}} &= (P_{gr}+P_{e})H_{\text{Ryd}}(P_{gr}+P_{e})
\\
&= \underbrace{P_{gr}H_{\text{Ryd}}P_{gr}}_{H_{gr}} + \underbrace{P_{e}H_{\text{Ryd}}P_{e}}_{H_{e}} + \underbrace{P_{e}H_{\text{Ryd}}P_{gr}}_{V_+} + \underbrace{P_{gr}H_{\text{Ryd}}P_{e}}_{V_-},
\end{split}
\end{equation}
where $V = V_+ + V_-$ couples the subspaces and the projection operator $P_{gr}= \mathds{1} - P_{e}$ is defined as $P_{gr} = \mathds{1}_{gr}^{(1)} \otimes \mathds{1}_{gr}^{(2)} \otimes \mathds{1}_{gr}^{(3)}$
with $\mathds{1}_{gr}^{(i)}\equiv \ket{g_i}\bra{g_i} + \ket{r_i}\bra{r_i}$.

The next step is to to diagonalize the Hamiltonian $H_{gr}$ in the form of $H_{gr}=\sum_l E_l P_l$, see Table~\ref{tab:H_gr}.
Let us to express $V_+$ in terms of the basis $P_l$ such that $V_+ =\sum_l V_+ P_l\equiv \sum_l V_+^l $. Following the procedure in Ref.~\cite{Reiter2012}, one can obtain an effective Hamiltonian and effective jump operators given by
\begin{align}
H_{\text{eff}} &= -\frac{1}{2} V_- \sum_l \Big[ (O_{\text{NH}}^l)^{-1} + \big( (O_{\text{NH}}^l)^{-1} \big)^\dagger \Big] V_+^l  + H_{gr},\\
c_{\text{eff}}^{(m)} &= c_m \sum_l (O_{\text{NH}}^{(l)})^{-1} V_+^l,
\end{align}
where $O_{\text{NH}}^{(l)}$ is a non-Hermitian operator which is defined as $O_{\text{NH}}^{(l)} = H_e - i/2 \sum_k c_{k}c_{k}^{\dagger} - E_l$, which in our case reads
\begin{equation*}
O_{\text{NH}}^{(l)} = \sum_{m=1}^{3}\big( \Delta_p - E_l - i \frac{\gamma_p}{2}\big) \ket{p_m}\bra{p_m} \otimes \mathds{1}_{gr} \otimes \mathds{1}_{gr} + V_{rr} \left(\ket{prr}\bra{prr}+ \ket{rrp}\bra{rrp}\right).
\end{equation*}
The effective process consists of an initial process $V_+^l$ from the state $l$, an intermediate propagation $O_{\text{NH}}$ in the dissipative subspace, and eventually either a coherent process $V_-$ to the reduced subspace corresponding to the effective Hamiltonian, or a decay to $\ket{g_m}$ representing the effective jump operators. In our case, the effective Hamiltonian and jump operators take the form
\begin{align}
H_{\text{eff}} &= V_{rr} \sum_{<ij>}\ket{r_i r_j}\bra{r_i r_j}
-\Omega_{\text{AUX}}^2 \sum_m  \frac{\Omega_{i\to f}}{\Omega_{i\to f}^2 + \gamma_p^2/4} \ket{r_m}\bra{r_m},\\
c_{\text{eff}}^{(m)} &= \frac{\Omega_{\text{AUX}} \sqrt{\gamma_p}}{\Omega_{i\to f} - i \gamma_p/2} \ket{g_m}\bra{r_m},
\end{align}
where $\Omega_{i\to f} = \Delta_p + \left(\tilde{\delta}_{\ket{f}} - \tilde{\delta}_{\bra{i}} \right) V_{rr}$ is a prefactor depending on the underlying state transition $\ket{f}\bra{i}$ such that 
\begin{equation*}
\tilde{\delta}_{\bra{i}} =
\begin{cases}
2 \hspace*{3mm} \text{if} \hspace*{3mm}  \bra{i} = \bra{rrr},\\
1 \hspace*{3mm} \text{if} \hspace*{3mm}  \bra{i} \in \{\bra{grr}, \bra{rrg}\},\\
0 \hspace*{3mm}  \text{otherwise}.\\
\end{cases}
\hspace{1cm}\text{and}\hspace{1cm}
\tilde{\delta}_{\ket{f}} =
\begin{cases}
1 \hspace*{3mm} \text{if} \hspace*{3mm}  \ket{f} \in \{\ket{prr}, \ket{rrp}\},\\
0 \hspace*{3mm}  \text{otherwise},\\
\end{cases}
\end{equation*}

\begin{table}[t]
	\centering
	\begingroup
	\setlength{\tabcolsep}{8pt} 
	\renewcommand{\arraystretch}{1.8} 
	\begin{tabular}[c]{l|l|l}
		\hline\hline
		Eigenstates of $H_{gr}$ & Eigenvalues of $H_{gr}$ & Coupling operator $V_+$ \\
		\hline
		$P_0 = \ket{ggg}\bra{ggg}$ & $E_0 = 0$ & $V_+^0 = 0$ \\
		\hline
		$P_1 = \ket{rgg}\bra{rgg}$ & $E_1 = 0$ & $V_+^1 = \Omega_{\text{AUX}}\ket{pgg}\bra{rgg}$ \\
		\hline
		$P_2 = \ket{grg}\bra{grg}$ & $E_2 = 0$ & $V_+^2 = \Omega_{\text{AUX}}\ket{gpg}\bra{grg}$ \\
		\hline
		$P_3 = \ket{ggr}\bra{ggr}$ & $E_3 = 0$ & $V_+^3 = \Omega_{\text{AUX}}\ket{ggp}\bra{ggr}$ \\
		\hline
		$P_4 = \ket{rgr}\bra{rgr}$ & $E_4 = 0$ & $V_+^4 = \Omega_{\text{AUX}}\big(\ket{pgr}+\ket{rgp}\big)\bra{rgr}$ \\
		\hline
		$P_5 = \ket{rrg}\bra{rrg}$ & $E_5 = V_{rr}$ & $V_+^5 = \Omega_{\text{AUX}}\big(\ket{prg}+\ket{rpg}\big)\bra{rrg}$ \\
		\hline
		$P_6 = \ket{grr}\bra{grr}$ & $E_6 = V_{rr}$ & $V_+^6 =  \Omega_{\text{AUX}}\big(\ket{gpr}+\ket{grp}\big)\bra{grr}$ \\
		\hline
		$P_7 = \ket{rrr}\bra{rrr}$ & $E_7 = 2V_{rr}$ & $V_+^7 = \Omega_{\text{AUX}}\big(\ket{prr}+\ket{rpr}+\ket{rrp}\big)\bra{rrr}$ \\
		\hline\hline
	\end{tabular}
	\endgroup
	\caption{Diagonalization of the ground-state Hamiltonian $H_{gr}$ and the decomposition of the coupling operator $V_+$.}
	\label{tab:H_gr}
\end{table}

To be explicit, the effective jump operators are given by
\begin{equation*}
\begin{split}
c_{\text{eff}}^1 &= \Omega_{\text{AUX}} \sqrt{\gamma_p} \big[ \lambda_1 \ket{ggg}\bra{rgg} + \lambda_1 \ket{ggr}\bra{rgr} + \lambda_2 \ket{grg}\bra{rrg} + \lambda_2 \ket{grr}\bra{rrr} \big],\\
c_{\text{eff}}^2 &= \Omega_{\text{AUX}} \sqrt{\gamma_p} \big[ \lambda_1 \ket{ggg}\bra{grg} + \lambda_2 \ket{rgg}\bra{rrg} + \lambda_2 \ket{ggr}\bra{grr} + \lambda_3 \ket{rgr}\bra{rrr} \big],\\
c_{\text{eff}}^3 &= \Omega_{\text{AUX}} \sqrt{\gamma_p} \big[ \lambda_1 \ket{ggg}\bra{ggr} + \lambda_1 \ket{rgg}\bra{rgr} + \lambda_2 \ket{grg}\bra{grr} + \lambda_2 \ket{rrg}\bra{rrr} \big],
\end{split}
\end{equation*}
where
\begin{align*}
\lambda_1 &= ( \Delta_p - i {\gamma}/{2})^{-1},\\ 
\lambda_2 &= ( \Delta_p - V_{rr}-i {\gamma}/{2})^{-1},\\
\lambda_3 &= ( \Delta_p - 2V_{rr} - i {\gamma}/{2})^{-1}.
\end{align*}
In particular, we are interested in having a completely local jump operator, i.e. $c_{a}^{(i)} = \sqrt{\gamma} \ket{g_i}\bra{r_i}$ with a tunable decay rate $\gamma$. This is possible by having a large detuing to the state $\ket{p_i}$, i.e. $\Delta_p\gg V_{rr}$, resulting in all the $\lambda_i$ being identical. 
The complex coefficient in $c_{\text{eff}}^{(m)}$ can be interpreted as an effective decay rate that is given by
\begin{equation}
\gamma = | \braketqqq{g_m}{c_{\text{eff}}^{(m)}}{r_m}|^2 =  \frac{\Omega_{\text{AUX}}^2}{\Delta_p^2+\gamma_p^2/4} \gamma_p.
\end{equation}
For the experimental parameters given in the main text, we have
$V_{rr} = 2\pi \times 120\,\textrm{MHz}$, which for $\Delta_p =
2\pi\times 500\,\mathrm{MHz}$ results in $\Omega_\text{AUX} = 2\pi
\times 80\,\text{MHz}$, which is well achievable for a low-lying
Rydberg state like $\ket{29S_{1/2}}$.

Furthermore, we wish to note that we have assumed a two-photon
excitation scheme connecting the ground state to the Rydberg state, in
order to have a dipole-allowed transition between the Rydberg and the
intermediate state. However, it is also possible to work with $P$
Rydberg states following a single-photon excitation, by replacing
$\Omega_\text{AUX}$ by a two-photon process combining the last two
stages of three-photon Rydberg excitation schemes
\cite{Gurian2012,Entin2013}.

	\section{Variational treatment of long-range correlations}
	\subsection{Derivation of the variational cost function}
	\label{SM:derivation}

	Here, we extend the variational principle for dissipative many-body systems to the regime of long-range interactions and correlations. Regarding the variational ansatz, we only include two-body long-range correlations and ignore higher-order correlations. This is justified according to the Bogoliubov-Born-Green-Kirkwood-Yvon hierarchy for correlation operators \cite{Navez2010}, asserting that in dissipative dynamics higher-order correlations decay faster than two-body terms. The variational ansatz is given by
	\begin{equation}
	\rho_{\text{var}}=\prod_{i=1}^N \rho_i+\sum_{i<j}\mathcal{R}C_{ij},
	\end{equation}
	where $C_{ij} = \rho_{ij}-\rho_i\otimes\rho_j$ denotes long-range correlations. For spin-1/2 systems, we can express the generic form of $\rho_i$ and $C_{ij}$ in terms of the Pauli matrices

	\begin{subequations}
		\begin{align}
		\rho_{i} &=\frac{1}{2} \Big(\mathds{1}_i+\sum_{\mu\in\{x,y,z\}}\alpha_{\mu} \sigma_{\mu}^{(i)}\Big),\\
		C_{ij} &=\frac{1}{4} \sum_{\substack{\mu,\nu\in \\ \{x,y,z\}}}  \zeta_{ij}(\mu, \nu) \sigma_{\mu}^{(i)} \sigma_{\nu}^{(j)}.
		\end{align}
	\end{subequations}
	where $\alpha_{\mu}$ denote the variational parameters and  $ \zeta_{ij}(\mu, \nu)$ represent variational correlation functions. Importantly, one has to define variational functions such that they preserve long-range effects of correlations but while keeping the number of variational parameters as low as possible in order to reduce numerical complexity of minimization. To this end, the variational correlation functions can be chosen based on prior knowledge about the stationary properties of the model we wish to analyze.
	
    Regarding the variational cost function, we base our definition according to the Hilbert-Schmidt norm that reads as
	\begin{equation}
	F_v \equiv\frac{\norm{\mathcal{L}\rho_{\text{var}}}_\text{HS}^2}{N \norm{\rho_{\text{var}}}_\text{HS}^2},
	\end{equation}
	which is normalised by the system size $N$ to have an intensive quantity and also the totally purity $\norm{\rho_{\text{var}}}_\text{HS}^2$ to makes the Hilbert-Schmidt norm unbiased towards any certain class of quantum states \cite{Kazemi2021b}.
	To find an efficiently computable variational norm, we expand the Liouvillian in terms of local and interacting terms, i.e. $\mathcal{L}=\sum_i \mathcal{L}_{i}+\frac{1}{2}\sum_{i\neq j} \mathcal{L}_{ij}$, where in the latter for the sake of simplicity we assume binary interactions. 
	For the numerator of $F_v$, we exploit the inner-product representation of the HS norm, i.e.
	\begin{equation}
	\begin{split}
	\norm{\mathcal{L}\rho}_\text{HS}^2 &=
	\braketvpp{\rho_{\text{var}}}{\mathcal{L}^\dagger \mathcal{L}}\\&=\braketvpp{\rho_{\text{var}}}{\sum_{i\neq j}\sum_{k\neq l}(\mathcal{L}_{i}+\frac{1}{2}\mathcal{L}_{ij})^\dagger (\mathcal{L}_{k}+\frac{1}{2}\mathcal{L}_{kl})}\\
	&\approx N\sum_{1\neq j\neq k\neq l}G_p^{\cancel{1jkl}}\braketvpp{\rho_{\text{var}}^{(1jkl)}}{\mathcal{L}_{1}^\dagger \mathcal{L}_{1}+\mathcal{L}_{1}^\dagger \mathcal{L}_{j}+2\mathcal{L}_{1}^\dagger \mathcal{L}_{1j}+\frac{1}{2}\mathcal{L}_{1j}^\dagger \mathcal{L}_{1j}\\&\hspace{3.3cm}+\mathcal{L}_{1}^\dagger \mathcal{L}_{jk}+\frac{1}{2}\mathcal{L}_{1j}^\dagger \mathcal{L}_{jk}+\frac{1}{2}\mathcal{L}_{1j}^\dagger \mathcal{L}_{1k}+\frac{1}{4}\mathcal{L}_{1j}^\dagger \mathcal{L}_{kl}}
	\end{split}
	\label{eq:crude_norm}
	\end{equation}
	where $G_p^{\cancel{ijkl}} =\braketv{\rho_{\text{var}}^{\cancel{ijkl}}}$ is the purity of the given density matrix corresponding to $N-4$ sites excluding sites $\{i, j, k, l\}$ and 
	\begin{equation}
	g_p=\braketv{\rho_1}=(1+\alpha_x^2+\alpha_y^2+\alpha_z^2)/2,
	\end{equation}
	is the single-site purity. The approximation in Eq.~\eqref{eq:crude_norm} comes from the exclusion of the marginal correlations, i.e. we assume $\rho_{\text{var}}^{(ijklm)}=\rho_{\text{var}}^{(ijkl)}\otimes\rho_{\text{var}}^{(m)}$. We have also made the assumption that the underlying system exhibits translational symmetry.  
	
	Finally, the variational cost function can be expressed as in the following form,

	\begin{equation}
	\begin{split}
	F_v =G_p^{-1}\biggl[&\sum_{1\neq j\neq k\neq l}G_p^{\cancel{1jkl}}\braketvpp{\rho_{\text{var}}^{(1jkl)}}{\mathcal{L}_{1}^\dagger \mathcal{L}_{1}+\mathcal{L}_{1}^\dagger \mathcal{L}_{j}+2\mathcal{L}_{1}^\dagger \mathcal{L}_{1j}+\frac{1}{2}\mathcal{L}_{1j}^\dagger \mathcal{L}_{1j}\\&\hspace{3.3cm}+\mathcal{L}_{1}^\dagger \mathcal{L}_{jk}+\frac{1}{2}\mathcal{L}_{1j}^\dagger \mathcal{L}_{jk}+\frac{1}{2}\mathcal{L}_{1j}^\dagger \mathcal{L}_{1k}+\frac{1}{4}\mathcal{L}_{1j}^\dagger \mathcal{L}_{kl}}\biggr].
	\end{split}
	\end{equation}
	This equation is still computationally very intensive even for small lattices. Therefore, we find a more efficient function which is an upper bound to the variational cost function, 
	\begin{equation}
	\begin{split}
	F_v \leq f_v&=
	g_p^{-2}\sum_{1\neq j}\braketvpp{\rho_{\text{var}}^{(1j)}}{\mathcal{L}_{1}^\dagger \mathcal{L}_{1}+\mathcal{L}_{1}^\dagger \mathcal{L}_{j}}\\
	&+g_p^{-3}\sum_{1\neq j\neq k}\braketvpp{\rho_{\text{var}}^{(1jk)}}{2\mathcal{L}_{1}^\dagger \mathcal{L}_{1j}+\mathcal{L}_{1}^\dagger \mathcal{L}_{jk}}\\
	&+g_p^{-4}\sum_{1\neq j\neq k\neq l}\braketvpp{\rho_{\text{var}}^{(1jkl)}}{\frac{1}{2}\mathcal{L}_{1j}^\dagger \mathcal{L}_{1j}+\frac{1}{2}\mathcal{L}_{1j}^\dagger \mathcal{L}_{jk}+\frac{1}{2}\mathcal{L}_{1j}^\dagger \mathcal{L}_{1k}+\frac{1}{4}\mathcal{L}_{1j}^\dagger \mathcal{L}_{kl}},
	\end{split}
	\end{equation}
	where here we made the approximation that $G_p(N)\approx G_p(N-n)g_p^n$ which is valid for $N\gg n$. Here, $n$ denotes the dimension of the variational states $\rho_{\text{var}}$ appearing in the variational norm.
	
	\subsection{Simplifying the upper bound}

        In the following, we will present further simplifications reducing the number of variational parameters in two different ways, by incorporating prior knowledge of the details of the long-range interacting Rydberg model. First, we introduce a pair correlation function to capture the correlations between different Rydberg excitation, which has a form that follows from prior knowledge from Rydberg-blockaded systems. Second, we employ the Heisenberg equations of motion to remove some of the variational parameters.

\subsubsection{Variational correlation functions}
The Rydberg blockade manifests itself in the two-body correlation among Rydberg states qualified by the well-known pair correlation function $g_2$ which reads  
\begin{equation}
g_2(r)=\frac{\sum_{i\neq j}g_2(i,j)\delta(r_{ij}-r)}{\sum_{i\neq j}\delta(r_{ij}-r)},
\end{equation}
with
\begin{equation}
g_2(i,j)=\frac{\langle P_r^{(i)} P_r^{(j)}\rangle}{{\langle P_r^{(i)} \rangle\langle P_r^{(j)}\rangle}}.
\end{equation}
The $g_2$ function is strongly suppressed within the blockade volume where no simultaneous Rydberg excitation can take place, i.e. $\langle P_r^{(i)} P_r^{(j)}\rangle\sim r^6$ for small values of $r$. On the other hand, for long enough distances the (Rydberg density-density) correlation vanishes, i.e. $\langle P_r^{(i)} P_r^{(j)}\rangle-{\langle P_r^{(i)} \rangle\langle P_r^{(j)}\rangle}=0$, and $g_2=1$. Due to the strong van der Waals interaction, this function exhibits a sharp blockaded-uncorrelated transition which roughly resembles a step function with some overshooting \cite{Loew2012}. Our ansatz for the $g_2$ function is
\begin{equation}
g_2(r) \approx\frac{\tilde{r}^6}{\tilde{r}^6+\alpha_{zz}^{(1)} + \alpha_{zz}^{(2)} \tilde{r} + \alpha_{zz}^{(3)} \tilde{r}^2},
\end{equation}
where $\tilde{r}\equiv\alpha_{zz}^{(4)} r$ in which $\alpha$'s are the
corresponding variational parameters. This function satisfies both
limits for small and large values of $r$. Using this simple function,
we can estimate the long-range correlations in driven-dissipative
Rydberg gases while keeping the number of variational parameters as
low as possible. In particular, the parameter $\alpha_{zz}^{(4)}$
controls the value of the blockade radius, while the remaining
parameters account for the details of the shape.

Now, one can easily express the $zz$ correlation function, i.e. $\zeta_{ij}(z,z)\equiv\langle\sigma_z^{(i)}\sigma_z^{(j)}\rangle-\langle\sigma_z^{(i)}\rangle\langle\sigma_z^{(j)}\rangle$, by expressing the definition of $g_2(r_{ij})$ in terms of $\sigma_z$. Crucially, we exploit translational invariance of the system, i.e., $\alpha_z = \langle\sigma_z^{(i)}\rangle \simeq \langle\sigma_z^{(j)}\rangle$. As a result, our variational ansatz for $zz$ correlation function is given by
\begin{equation}
\zeta_{ij}(z,z) = [g_2(r_{ij}) -1] (\alpha_z+1)^2.
\end{equation}

Using the Heisenberg equations of motion we can further decrease the number of variational parameters. In the case of the dissipative Rydberg model, one can find closed relations between some variational parameters by solving the corresponding Heisenberg equations for the steady state solution,
\begin{equation}\label{eq:Ryd_Heisenberg}
\begin{split}
&\frac{d}{dt}\langle\sigma_z^{(i)}\rangle=0 \rightarrow  \alpha_y = \frac{\gamma_a}{\Omega} (\alpha_z+1),\\ 
&\frac{d}{dt}\langle\sigma_z^{(i)}\sigma_z^{(j)}\rangle=0 \rightarrow  \zeta_{ij}(y,z) = \frac{\gamma_a+\gamma_p}{\Omega} \zeta_{ij}(z,z),\\ 
&\frac{d}{dt}\langle\sigma_x^{(i)}\rangle=0 \rightarrow  \alpha_x = \frac{\gamma_a(\alpha_z+1)}{\Omega(\gamma_a+4\gamma_p)} \left[2\Delta-C_6(\alpha_z+1)\sum_{j\neq i}\frac{g_2(r_{ij})}{r_{ij}^6}\right],
\end{split}
\end{equation}

For the remaining correlation functions, we use a simple exponential decay, which is in good agreement with numerical simulations of small systems. In Table~\ref{tab:Var_Funs} all the aforementioned ans{\"a}tze are presented. Therefore, in total we have formulated the variational many-body state in terms of 13 parameters.

\begin{table}[t]
	\centering
	\begingroup
	\setlength{\tabcolsep}{8pt} 
	\renewcommand{\arraystretch}{1.8} 
	\begin{tabular}[c]{c|c|c}
		\hline\hline
		\multirow{2}{5em}{correlation function} & \multirow{2}{5em}{variational parameters} & \multirow{2}{5em}{variational ansatz} \\
		&&\\
		\hline
		$\zeta_{ij}(x,x)$ &$\alpha_{xx}^{(1)}$, $\alpha_{xx}^{(1)}$& ${\alpha_{xx}^{(2)} \exp[-\alpha_{xx}^{(2)}r_{ij}]}$ \\[1ex] 
		\hline
		$\zeta_{ij}(y,y)$&$\alpha_{yy}^{(1)}$, $\alpha_{yy}^{(1)}$& ${\alpha_{yy}^{(2)} \exp[-\alpha_{yy}^{(2)}r_{ij}]}$ \\ 
		\hline
		$\zeta_{ij}(z,z)$&$\alpha_{z}$, $\alpha_{zz}^{(1)}$, $\alpha_{zz}^{(2)}$, $\alpha_{zz}^{(3)}$, $\alpha_{zz}^{(4)}$& $[g_2(r) -1] (\alpha_z+1)^2$ \\
		\hline
		$\zeta_{ij}(x,y)$&$\alpha_{xy}^{(1)}$, $\alpha_{xy}^{(2)}$& ${\alpha_{xy}^{(1)} \exp[-\alpha_{xy}^{(2)}r_{ij}]}$ \\
		\hline
		$\zeta_{ij}(x,z)$&$\alpha_{xz}^{(1)}$, $\alpha_{xz}^{(2)}$& ${\alpha_{xz}^{(1)} \exp[-\alpha_{xz}^{(2)}r_{ij}]}$ \\
		\hline
		$\zeta_{ij}(y,z)$ &$\alpha_{z}$, $\alpha_{zz}^{(2)}$, $\alpha_{zz}^{(2)}$, $\alpha_{zz}^{(3)}$, $\alpha_{zz}^{(4)}$& $(\gamma_a+\gamma_p)[g_2(r) -1] (\alpha_z+1)^2/\Omega$ \\
		\hline\hline
	\end{tabular}
	\endgroup
	\caption{Varitional ansatz for two-body correlation functions in two-dimensional driven-dissipative Rydberg gases.\label{tab:Var_Funs}}
\end{table}

It is worth mentioning that our variational ansatz is mostly expanded around the diagonal elements of the density matrix. This is because all the variational correlation functions are exponentially decaying with the distance except the $zz$ and $yz$ correlation functions that only contribute to diagonal and semi-diagonal elements respectively (see Table~\ref{tab:Var_Funs}). This is consistent with the reported Monte-Carlo simulation of the dynamics showing that the dissipative processes destroy quantum correlations while enhancing classical correlations \cite{Petrosyan2013}. As a result, the long-time limit of the total density matrix is nearly diagonal. 

\subsubsection{Minimization constraints}

Having fixed the variational manifold, we have to choose a proper minimization method to be able to find the optimal solution for the associated steady state. In this specific model, since the variational landscape is relatively benign, we make use of a fast-converging local non-linear optimization method based on the gradient-descent scheme. In particular, we use SLSQP Optimization subroutine of SciPy library in Python \cite{Virtanen2020}. The implemented SLSQP method supports constrained optimization which is necessary for the variational method as discussed below.

Specifically, the minimization of the variational norm is constrained
by two inequalities. The first inequality ensures the positivity of
the quantum state. Additionally, we impose a second inequality
constraint derived from the ansatz for the pair correlation
function. Explicitly, these inequalities are given by
\begin{enumerate}
	\item Positivity of the density matrix $\rho$: As the variational norm consists of bipartite reduced density matrices, we only have positivity constraints on all these matrices
	\begin{equation*}
	\braketp{\rho_{ij}}{\Psi} \geqslant 0 \hspace*{3mm} \text{for any} \hspace*{2mm} \ket{\Psi},
	\end{equation*}	
	that $\rho_{ij}=\text{Tr}_{\cancel{ij}}[\rho]$. From numerical point of view, this constraint is equivalent to the positivity of the smallest eigenvalue of $\rho_{ij}$,
	\begin{equation*}
	\text{E}_1[\rho_{ij}] \geqslant 0 \hspace{5mm} \text{for all} \hspace{2mm} \rho_{ij}.
	\end{equation*}			
	
	\item Normalization condition: In equilibrium, normalization condition of the pair correlation function $g_2(\mathbf{r})$ leads to $\int d\mathbf{r}[1-g_2(\mathbf{r})] =\braket{P_r}^{-1}$ \cite{Weimer2008a}. In the non-equilibrium case, this equality condition turns into the following inequality constraint
	\begin{equation*}
	\sum_{i\neq j} r_{ij}[1-g_2(\mathbf{r})] \leqslant \frac{1}{\braket{P_r}}.
	\end{equation*}
\end{enumerate}

\section{Effective short-range description of frozen Rydberg gases}
	\label{SM:short-range}
The Rydberg Hamiltonian resembles the transverse-field Ising Hamiltonian with long-range interactions, i.e.
\begin{equation}
H_{\text{Ising}} = \frac{\Omega}{2} \sum_{i}{\sigma_x^{(i)}} + V \sum_{i<j}{\frac{\sigma_z^{(i)}\sigma_z^{(j)}}{|\textbf{r}_i-\textbf{r}_j|^6}}.
\end{equation}
However, there are two differences between these two Hamiltonians. First, the Ising interaction $\sigma_z^{(i)}\sigma_z^{(j)}$  does not exclude ground state interactions as the Rydberg-Rydberg interaction does. Second, the Rydberg Hamiltonian contains additional terms proportional to $\sigma_z^{(i)}$ which corresponds to a longitudinal magnetic field in the spin language. However, by expanding Rydberg interactions in terms of $\sigma_z$, one can obtain Ising interactions and cancel out the longitudinal field by an effective detuning. Nevertheless, reducing Rydberg interactions to Ising ones may destroy the particular properties of Rydberg gases such as the blockade effect. Hence, we effectively mimic the blockade by discarding sites up to a truncation radius $r_d$ and keeping the subsequent long-range tail in an Ising form. This approach can be seen as a long-range extension of so-called ``perfect blockade'' models \cite{Sun2008,Olmos2010}. Then, we obtain
\begin{equation}
\begin{split}
H_{\text{Ryd}} =& -\frac{\Delta}{2} \sum_{i}{\sigma_z^{(i)}} + \frac{\Omega}{2} \sum_{i}{\sigma_x^{(i)}} + \frac{C_6}{a^6} \sum_{r_{ij}\leq r_d }{\frac{P_r^{(i)}P_r^{(j)}}{r_{ij}^6}}\\
& + \frac{C_6}{4a^6} \sum_{r_{ij}> r_d }{\frac{\sigma_z^{(i)}+\sigma_z^{(j)}}{r_{ij}^6}}+ \frac{C_6}{4a^6} \sum_{r_{ij}> r_d }{\frac{\sigma_z^{(i)}\sigma_z^{(j)}}{r_{ij}^6}}+ \frac{C_6}{4a^6} \sum_{r_{ij}> r_d }{\frac{\mathds{1}^{(i)}\mathds{1}^{(j)}}{r_{ij}^6}},
\end{split}
\end{equation}
where $r_{ij}=|\textbf{r}_i-\textbf{r}_j|$. Since the last term only shifts the total energy of the system, we can ignore it. Hence, we have
\begin{equation}
\begin{split}
H_{\text{Ryd}} =& \Big(-\frac{\Delta}{2}+V_z\Big) \sum_{i}{\sigma_z^{(i)}} + \frac{\Omega}{2} \sum_{i}{\sigma_x^{(i)}} \\&+ \frac{C_6}{a^6} \sum_{r_{ij}\leq r_d }{\frac{P_r^{(i)}P_r^{(j)}}{r_{ij}^6}}+ \frac{C_6}{4a^6} \sum_{r_{ij}> r_d }{\frac{\sigma_z^{(i)}\sigma_z^{(j)}}{r_{ij}^6}},
\end{split}
\end{equation}
where
\begin{equation}
V_z=\frac{C_6}{4a^6} \sum_{r_{ij}> r_d }{\frac{1}{r_{ij}^6}},
\end{equation}
represents the interaction-induced shift in the longitudinal magnetic field. 
 
As mentioned before, we can cancel out the longitudinal field by an appropriate choice of the detuning, i.e., $\Delta(r_d) = 2V_z(r_d)$. Crucially, at the extreme case of $r_d=0$, the resulting value of detuning $\Delta(r_d=0)$ is very large such that it compensates for the energy shift by the strong vdW interaction, leading to the anti-blockade regime. However, we are interested in the regime of finite $r_d$ where the Rydberg blockade plays an important role. In the following, we refer to these values of detuning as transverse points. To quantify these points, let's first evaluate the $V_z$ which is related to $r_d$ as following
\begin{equation}
\label{V_Ising}
V_z(r_d)=\frac{C_6}{4a^6} \Bigg[\sideset{}{'}\sum_{r_{ij}=-\infty}^{+\infty}{\frac{1}{r_{ij}^6}} - \sum_{r_{ij}\leq r_d }{\frac{1}{r_{ij}^6}} \Bigg],
\end{equation}
where the primed sum indicates that the zero point is excluded. Using the notion of Dirichlet series, the first term of the above equation describing distances on a square lattice, is equivalent to
\begin{equation}
\begin{split}
\sideset{}{'}\sum_{-\infty}^{+\infty}{\frac{1}{r_{ij}^6}}&= \sideset{}{'}\sum_{-\infty}^{+\infty}{\frac{1}{(i^2+j^2)^3}}\\
& = 4 \zeta (3) \beta (3),
\end{split}
\end{equation}
where $\beta(x)$ is the Dirichlet $\beta$-function and $\zeta(x)$ is the Riemann  $\zeta$-function \cite{Actor1990}. We recall that $\beta(3)=\pi^3/32$ and $\zeta(3)\approx1.202057$ is the sum of the reciprocals of the positive cubes which is known as the Apery's constant. Therefore, we can express Eq.~\eqref{V_Ising} as follows
\begin{equation}
V_z=\frac{C_6}{a^6} \Bigg[\zeta (3)\beta (3) - \sum_{r_{ij}\leq r_d }{\frac{f(i,j)}{r_{ij}^6}} \Bigg],
\end{equation}
where
\begin{equation*}
f(i,j) = 
\begin{cases}
2 \hspace*{3mm} \text{if} \hspace*{3mm} i \neq 0 \hspace*{1mm} \text{and} \hspace*{1mm} i \neq j,\\
1 \hspace*{3mm} \text{otherwise}.
\end{cases}
\end{equation*}
In Table~\ref{tab:transpoints}, we present the transverse points of the laser detuning for $0<r_{d}<3$ and the nearest-neighbor interaction strength of $C_6/a^6\approx h\times 882\text{MHz}$.
\begin{table}[t]
	\centering
	\begingroup
	\setlength{\tabcolsep}{8pt} 
	\renewcommand{\arraystretch}{1.8} 
	\begin{tabular}[c]{c|c|c}
		\hline\hline
		$r_d$ & $V_za^6/C_6 $ & $\Delta/2\pi [\text{MHz}]$ \\
		\hline
		$0$ & $ \zeta (3)\beta (3) $ & $2055.0144$ \\
		\hline
		$1$ & $ \zeta (3)\beta (3)-\left(1\right) $ & $290.6422$ \\
		\hline
		$\sqrt{2}$ & $\zeta (3)\beta (3)-\left(1+\frac{1}{8}\right)$ & $70.0957$ \\
		\hline
		$2$& $\zeta (3)\beta (3)-\left(1+\frac{1}{8}+\frac{1}{64}\right)$ & $42.5274$ \\
		\hline
		$\sqrt{5}$& $\zeta (3)\beta (3)-\left(1+\frac{1}{8}+\frac{1}{64}+\frac{2}{125}\right)$ & $14.2974$ \\
		\hline
		$\sqrt{8}$& $\zeta (3)\beta (3)-\left(1+\frac{1}{8}+\frac{1}{64}+\frac{2}{125}+\frac{1}{512}\right)$ & $ 10.8514$ \\
		\hline
		$3$&  $\zeta (3)\beta (3)-\left(1+\frac{1}{8}+\frac{1}{64}+\frac{2}{125}+\frac{1}{512}+\frac{1}{729}\right)$ & $8.4311$ \\ 
		\hline\hline
	\end{tabular}
	\endgroup
	\caption{Transverse points of the laser detuning in the Rydberg Hamiltonian for $C_6=h\times 20 \text{MHz}\mu\text{m}^6$ and $a= 532\,\text{nm}$.}
	\label{tab:transpoints}
\end{table}

To  proceed further, it is possible to have an effective low-energy description of the aforementioned model with the associated truncation radius $r_d$. A short-range Ising model can be used as an effective model with the Hamiltonian,
\begin{equation}
H_{\text{eff}} = h_z(r_d) \sum_{i}{\sigma_z^{(i)}} + \frac{\Omega}{2} \sum_{i}{\sigma_x^{(i)}} + J_{\text{eff}}(r_d) \sum_{\braket{ij}}\sigma_z^{(i)}\sigma_z^{(j)},
\end{equation}
with $h_z(r_d)=-\Delta/2+V_z$ and $J_{\text{eff}}(r_d)=V_z$. This effective model is a good candidate for describing the original model at the transverse points, i.e. $h_z(r_d)=0$.